\documentclass[%
 reprint,
preprintnumbers,
nofootinbib,
amsmath,amssymb,aps,superscriptaddress
]{revtex4-2}
\counterwithout{equation}{section}
\usepackage{amsmath,float,graphicx,placeins,amssymb,multirow,pgfplots,pgfplotstable}
\usepackage{empheq}
\usepackage{booktabs}
\usepackage{tikz}
\usepackage{comment}
\usepackage{enumerate,slashed,cancel,comment,color,bbm,graphicx,rotating,placeins,mathtools,multirow,hhline}
\usepackage[normalem]{ulem}
\usepackage{array}
\usepackage{hyperref}
\usepackage{color}
\usepackage{booktabs}
 \hypersetup
 {
   colorlinks,
   linkcolor={blue!80!black},
   citecolor={blue!70!black},
   urlcolor={blue!70!black}
 }

\usepackage{hyperref}
\usepackage{tabularx}

\newcommand{\be}{\begin{equation}}
\newcommand{\ee}{\end{equation}}
\newcommand{\bea}{\begin{eqnarray}}
\newcommand{\eea}{\end{eqnarray}}
\newcommand{\ba}{\begin{aligned}}
\newcommand{\ea}{\end{aligned}}

\newcommand{\Planck}{{\it Planck}}
\newcommand{\ACT}{{ACT }}

\pgfplotsset{compat=1.18} 
\begin{document}

\title{
Running of the spectral index: Reconciling the CMB with the Lyman-$\alpha$ Forest
}

\begin{abstract} 

We investigate the scale dependence of the primordial power spectrum by combining \Planck, ACT DR6, SPT-3G, and eBOSS Lyman-$\alpha$ forest data, extending sensitivity to smaller comoving scales than those probed by the CMB alone. Within a parametrisation based on a Taylor expansion around the pivot scale, we constrain the running of the spectral index $\alpha_s$ and its running $\beta_s$.
By using eBOSS likelihoods exhibiting a suppression of small-scale power either in amplitude or spectral index, we show that the latter can be accommodated by correlated variations of $(\alpha_s,\beta_s)$, leading to a preference for non-zero running. We show that inflationary potentials with localised features—such as Gaussian dips, bumps, or axion-monodromy modulations—can reproduce the inferred scale dependence while remaining compatible with current CMB constraints.
We release the public {\tt PIPE} code to enable systematic tests of inflationary potentials against current CMB datasets.
\end{abstract}

\author{Malcolm Fairbairn}
\email[Email address: ]{malcolm.fairbairn@kcl.ac.uk}

\author{Lucien Heurtier}
\email[Email address: ]{lucien.heurtier@kcl.ac.uk}

\author{Mar\'ia Olalla Olea-Romacho}
\email[Email address: ]{maria\_olalla.olea\_romacho@kcl.ac.uk}

\affiliation{Theoretical Particle Physics and Cosmology, King’s College London,\\ Strand, London WC2R 2LS, United Kingdom}

\maketitle

\section{Introduction}

Cosmic inflation provides the leading explanation for the initial conditions of our Universe, accounting for its homogeneity and flatness while generating the primordial fluctuations observed in the cosmic microwave background (CMB) and large-scale structure (LSS) \cite{Liddle:1993fq,Baumann:2022mni}. Yet, the physical origin of this accelerated expansion---and the detailed shape of the inflaton potential that may drive it---remain unknown.

\begin{figure*}
    \centering
    \includegraphics[width=0.5\linewidth]{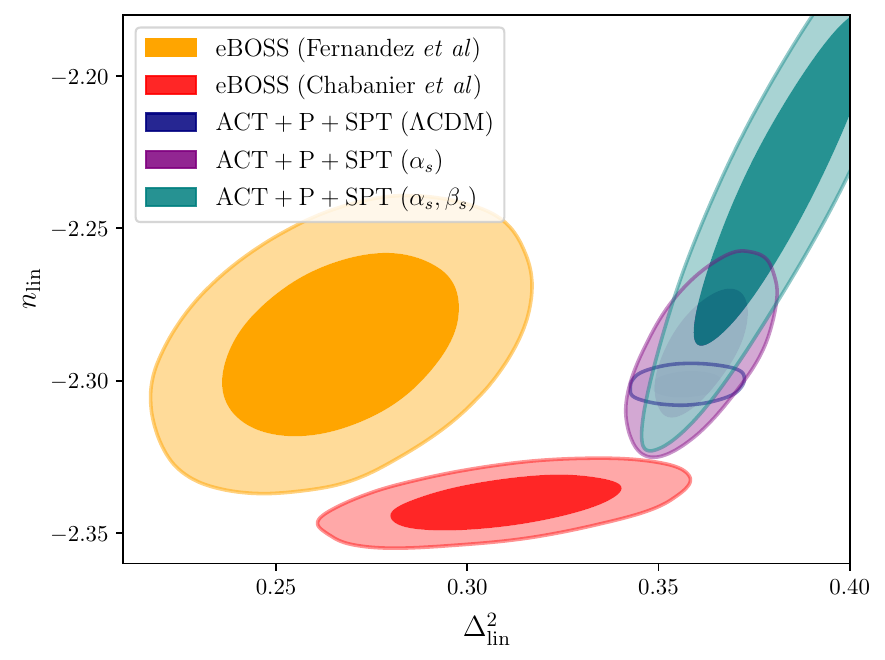}
    \caption{\footnotesize Posteriors for $(\Delta_{\rm lin}^2,n_{\rm lin})$ when comparing ACT+P (left panel) and ACT+P+SPT (right panel) to the eBOSS likelihoods from Chabanier {\it et al}~\cite{eBOSS:2018qyj} and Fernandez {\it et al}~\cite{Fernandez:2023grg}, in the $\Lambda$CDM model (blue contours), and when including the running (purple contours) and running of the running (teal contours).}
    \label{fig:nlin}
\end{figure*}

Since its original proposal~\cite {Starobinsky:1979ty, Guth:1980zm}, CMB observatories and LSS surveys have enabled increasingly precise tests of inflation. The beautiful successes of {\it WMAP}~\cite{WMAP:2010qai} and \Planck~\cite{Planck:2018vyg} revealed the structure of the CMB fluctuations on small scales, providing evidence for a Universe fully consistent with the $\Lambda$CDM hypothesis -- a spatially flat Universe containing only baryons, cold dark matter and a cosmological constant, with a flat or nearly flat dimensionless power spectrum of scalar curvature perturbations that can be modelled with a single power law $\Delta^2_{\mathcal R, \Lambda\mathrm{CDM}}(k) \equiv A_s ({k}/{k_\star})^{n_s-1}$. Cosmic inflation typically predicts a spectral index $n_s$ close to unity, indeed, \Planck~measured $n_s = 0.9651 \pm 0.0044$ at the  pivot scale $k_\star = 0.05 \, {\rm Mpc}^{-1}$ \cite{Planck:2018vyg}, confirming the success of the $\Lambda$CDM model. In single-field slow-roll inflation, the dynamics of the inflaton---uniquely determined by the shape of the inflaton potential---predict small but generic deviations from this form of the primordial power spectrum (PPS), which can be captured around a pivot scale by Taylor-expanding its logarithm as
\bea\label{eq:running}
\ln\Delta^2_{\mathcal R}(k) &\equiv& \ln A_s +(n_s-1) \ln\left(\frac{k}{k_\star}\right)\nonumber\\
&+&\frac{\alpha_s}{2} \ln\left(\frac{k}{k_\star}\right)^2 + \frac{\beta_s}{6} \ln\left(\frac{k}{k_\star}\right)^3\nonumber\\
&+& \hdots
\eea
where $\alpha_s$ and $\beta_s$ denote the running and running-of-the-running of the spectral index.

While CMB observations tightly constrain this spectrum around the pivot scale, their sensitivity decreases at smaller scales, limiting their ability to detect mild departures from a power-law spectrum. To probe these smaller scales, we incorporate data from eBOSS~\cite{eBOSS:2018qyj}, which measured $\sim 2 \times 10^5$ quasar spectra with detectable Lyman-$\alpha$ forests.\footnote{Shortly after the completion of this work, a new Lyman-$\alpha$ forest analysis based on DESI DR1 data became available~\cite{Chaves-Montero:2026hqd}, reporting constraints consistent with a primordial spectrum with negligible running. The eBOSS likelihood of Chabanier \textit{et al.}~(2019)~\cite{eBOSS:2018qyj} adopted here does not incorporate the latest modelling improvements in Lyman-$\alpha$ analyses, and should therefore be regarded as a benchmark scenario exhibiting a suppression of small-scale power. In this sense, our results are best interpreted as characterising how such a suppression would map onto primordial parameters and inflationary dynamics, rather than as evidence for its presence in current data. Future measurements with improved precision and control of systematics will be crucial to establish whether such scale-dependent features are supported.} These observations probe wavenumbers around $k \sim 1 \, h \, \mathrm{Mpc}^{-1}$, providing complementary information on the amplitude and slope of the matter power spectrum in a regime inaccessible to the CMB alone. 

Recent CMB measurements from ACT DR6~\cite{ACT:2025fju, ACT:2025tim} mildly favour a positive running, $\alpha_s > 0$. These experiments probe smaller angular scales than \textit{Planck}, extending CMB sensitivity to higher wavenumbers and thereby bridging the gap with the eBOSS Lyman-$\alpha$ forest regime. However, some analyses of the eBOSS data suggested a suppression of small-scale power relative to the nearly scale-invariant spectrum preferred by CMB measurements. In particular, Rogers and Poulin~\cite{Rogers:2023upm} reported a significant deviation from a single power-law primordial spectrum when combining \Planck\ with eBOSS data, which can be interpreted as a preference for negative running of the spectral index, therefore pointing in the opposite direction of recent CMB measurements. In this paper, we explore how such results evolve when accounting for new CMB datasets, using different alternative likelihoods~\cite{Fernandez:2023grg, Walther:2024tcj}, but also allowing for the running of the running $\beta_s$ of the spectral index, which was not present in previous analyses.

The likelihood~\cite{eBOSS:2018qyj} used in Ref.~\cite{Rogers:2023upm} does not reflect the most recent developments in the analysis of eBOSS Lyman-$\alpha$ data, for which updated methodologies have since become standard in the literature~\cite{Fernandez:2023grg, Walther:2024tcj}. In particular, as illustrated in Fig.~\ref{fig:nlin}, different eBOSS likelihoods, such as the ones derived in~\cite{eBOSS:2018qyj} and~\cite{Fernandez:2023grg}, compare fairly differently to CMB likelihoods, populating different regions in the $(\Delta^2_{\rm lin}, n_{\rm lin})$ plane, where $\Delta^2_{\rm lin} \equiv \frac{k^3 P_{\rm lin}(k_p, z_p)}{2\pi^2}$ and $n_{\rm lin} \equiv \left. \frac{{\rm d} \ln P_{\rm lin}(k, z)}{{\rm d} \ln k} \right|{k_p, z_p}$ denote the amplitude and local slope of the linear matter power spectrum $P_{\rm lin}(k, z)$, evaluated at $z_p = 3$ and $k_p = 0.009 \, \mathrm{s km^{-1}}$ (see Sec.~\ref{sec:powerspectrumrunning} for details). Indeed, whereas the likelihood~\cite{Fernandez:2023grg} favours a strong suppression of the spectrum amplitude only, the one from Ref.~\cite{eBOSS:2018qyj} points towards a milder suppression of the amplitude but a clear suppression of the tilt.

In this work, we assume that these deviations originate from primordial physics~\cite{Hannestad:2001nu}, and perform a joint analysis of \Planck, ACT DR6, SPT-3G~\cite{SPT-3G:2025bzu}, and eBOSS data using both the likelihoods from Ref.~\cite{eBOSS:2018qyj} and Ref.~\cite{Fernandez:2023grg}. We use these likelihoods as representative benchmarks of what could be potential suppressions in the matter power spectrum shape to study how this type of suppression can be translated into correlated variations of $\alpha_s$ and $\beta_s$, and to identify which inflationary features can be described within this framework. As we will see in the next sections, the Taylor-expanded parametrisation used in Eq.~\eqref{eq:running} only captures a restricted class of smooth distortions of the primordial power spectrum, and will appear to be better suited to describe deviations of the type associated with the likelihood of Ref.~\cite{eBOSS:2018qyj}. We then test representative scenarios—including monomial, axion-monodromy, and potentials with localised features—and determine which can reproduce the inferred small-scale behaviour while remaining consistent with CMB constraints. We also release the public {\tt PIPE} code to evaluate arbitrary inflationary potentials against current CMB data.

\begin{figure}
    \centering
    \includegraphics[width=0.92\linewidth]{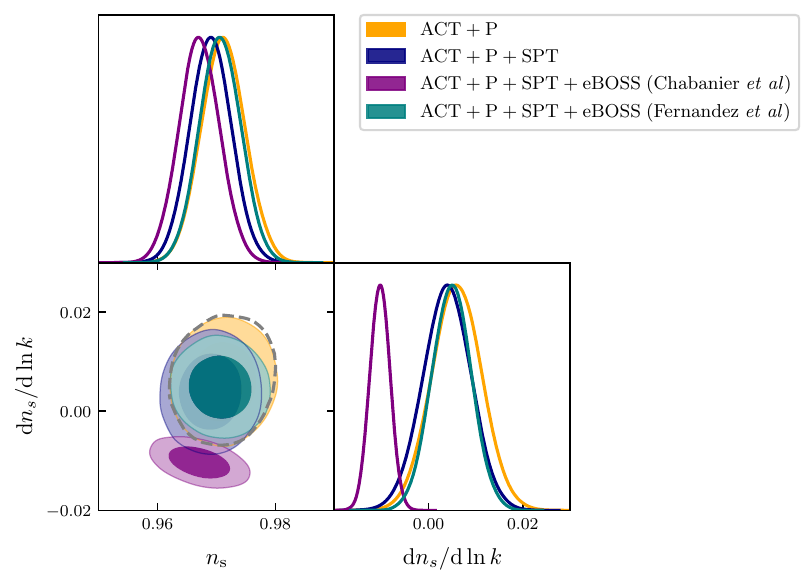}
    \caption{\footnotesize Joint and marginalized posterior probability distributions for $n_s$ and its running $\alpha_s = \mathrm{d}n_s/\mathrm{d}\ln k$, derived from the three likelihood combinations analysed in this work. Shaded contours represent the 68\% and 95\% credible regions. The dashed grey curve corresponds to the ACT+P constraint from Ref.~\cite{ACT:2025tim}, obtained when fitting only $\alpha_s$.
}
    \label{fig:corner_norunrun}
\end{figure}
\begin{figure}
    \includegraphics[width=\linewidth]{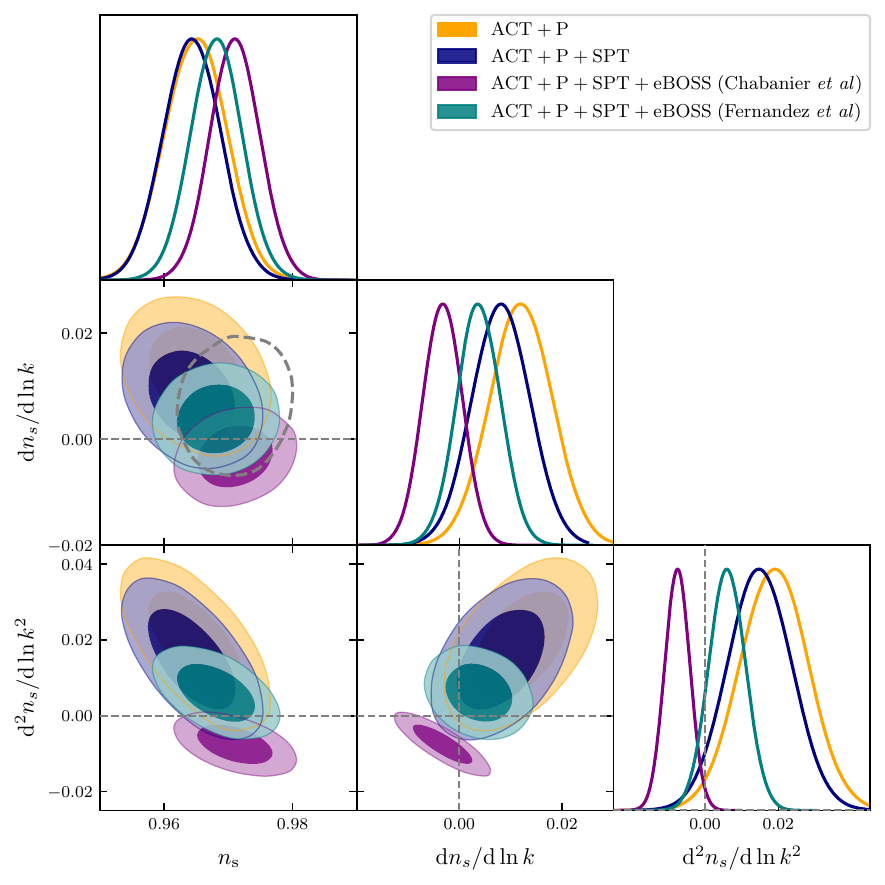}
    \caption{\footnotesize Same as FIG.~\ref{fig:corner_norunrun} but including $\beta_s = \mathrm{d}^2 n_s/\mathrm{d}\ln k^2$. Median values and corresponding 68\% credible intervals for these parameters are reported in the Supplemental Material.}
    \label{fig:corner_nrunrun}
\end{figure}

\section{Power Spectrum Running}
\label{sec:powerspectrumrunning}
\textit{Likelihood.} To constrain the PPS, we use the most recent likelihoods from \Planck~\cite{Planck:2018vyg} and the ACT DR6 Lite likelihood released by the ACT collaboration~\cite{ACT:2025fju,ACT:2025tim}.\footnote{\url{https://github.com/ACTCollaboration/DR6-ACT-lite}} Our setup follows the \Planck-ACT ({P-ACT}) cut configuration described in Ref.~\cite{ACT:2025fju}, which combines large-scale \Planck~data with high-resolution ACT measurements. Specifically, we include the low-$\ell$ TT spectrum from \Planck~\cite{Planck:2019nip} and the low-$\ell$ EE spectrum from the \texttt{Sroll2} reanalysis~\cite{Pagano:2019tci}, while at high multipoles we use ACT DR6 CMB-only data in the \Planck-\ACT cut, retaining \Planck~data only up to $\ell = 1000$ in TT and $\ell = 600$ in TE/EE. ACT data dominate at smaller scales, covering the range $600 < \ell < 6500$ in TT, TE, and EE, with foregrounds marginalised over. We also include the foreground-marginalised likelihood \texttt{SPT-3G D1 Lite} published recently by SPT~\cite{SPT-3G:2025bzu}.\footnote{\url{https://github.com/SouthPoleTelescope/spt_candl_data}} 
Similarly to Ref.~\cite{Rogers:2023upm}, we incorporate Lyman-$\alpha$ forest constraints on the linear matter power spectrum using a compressed two-dimensional Gaussian likelihood constructed from the eBOSS (Sloan Digital Sky Survey Data Release~14; SDSS~DR14) Lyman-$\alpha$ forest flux power spectrum~\cite{eBOSS:2018qyj}. In this approach, the cosmological information contained in the flux power spectrum is projected onto the two effective parameters: the amplitude $\Delta^2_{\rm lin}$ and local spectral tilt $n_{\rm lin}$ of the linear matter power spectrum $P_{\rm lin}(k,z)$, evaluated at the pivot redshift $z_p = 3$ and wavenumber $k_p = 0.009\,\mathrm{s\,km^{-1}}$.
We first consider the compressed likelihood derived in Ref.~\cite{Goldstein:2023gnw}, which is based on the eBOSS DR14 Lyman-$\alpha$ analysis of Ref.~\cite{eBOSS:2018qyj} and based on the BOSS/eBOSS Lyman-$\alpha$ forest flux power spectrum measurements reported in Ref.~\cite{El-Ella:2019gnc}. In this case, the mean values of $\Delta^2_{\rm lin}$ and $n_{\rm lin}$ are $0.310$ and $-2.340$, with standard deviations $0.020$ and $0.006$, respectively, and a correlation coefficient $\rho = 0.512$~\cite{Goldstein:2023gnw}.

In addition, we also consider the alternative Lyman-$\alpha$ likelihood proposed in Ref.~\cite{Fernandez:2023grg}, which adopts a more conservative treatment of astrophysical and modelling uncertainties affecting the inference of the linear matter power spectrum. For consistency with the analysis described above, we also approximate this likelihood by a two-dimensional Gaussian in the $(\Delta^2_{\rm lin}, n_{\rm lin})$ plane. In this Gaussian approximation, the mean values are taken to be $\Delta_{\mathrm{lin}}^2 = 0.267^{+0.018}_{-0.023}$ and $n_{\mathrm{lin}} = -2.288^{+0.020}_{-0.020}$, with a correlation coefficient of $0.4$. These values are chosen to reproduce the central values, uncertainties, and degeneracy directions reported in Ref.~\cite{Fernandez:2023grg}.

\paragraph*{Monte Carlo Markov Chain.}
We perform Monte Carlo Markov Chain (MCMC) sampling using the publicly available code \texttt{Cobaya}~\cite{Torrado:2020dgo} interfaced with the Boltzmann solver \texttt{CAMB}~\cite{Lewis:1999bs}, to extract constraints on the amplitude and shape of the PPS given in Eq.~\eqref{eq:running}. We perform two separate analyses: one where we sample the scalar amplitude $\ln(10^{10} A_s)$ and spectral index $n_s$ at the pivot scale $k_{\star}=0.05 \, \rm{Mpc}^{-1}$, together with the running $\alpha_s$, and one in which we also include the running of the running $\beta_s$. These parameters are varied jointly with the remaining four baseline $\Lambda$CDM parameters: the optical depth to reionization $\tau$, the physical baryon density $\omega_b$, the physical cold dark matter density $\omega_c$, and the angular size of the sound horizon $\theta_s$. In addition, we vary the nuisance parameters $T_{\rm cal}$ and $E_{\rm cal}$, which account for the absolute calibration of the temperature and polarisation spectra, respectively. 
For all parameters, we adopt the same flat priors as used by the ACT collaboration in the ACT-DR6.02 ``nrun'' analysis.\footnote{\url{https://lambda.gsfc.nasa.gov/product/act/act_dr6.02/act_dr6.02_chains_nrun_get.html}} To ensure convergence, we run multiple chains (using 12 CPUs when including only $\alpha_s$ in the analysis, and 15 CPUs when also including $\beta_s$) in parallel and monitor the Gelman-Rubin $R$ statistic, requiring $R-1<0.05$ for all sampled parameters. 

\paragraph*{Results.}

Fig.~\ref{fig:nlin} shows the regions favoured by ACT+P+SPT for $\Lambda$CDM (blue), running ($\alpha_s$, purple), and running-of-the-running ($\alpha_s,\beta_s$, teal), compared to those preferred by eBOSS Lyman-$\alpha$ analyses. The red contours correspond to the likelihood of Chabanier {\it et al.}~\cite{eBOSS:2018qyj} used in this work, while the yellow contours show an alternative analysis based on the \texttt{PRIYA} simulations~\cite{Fernandez:2023grg}, which rely on a more up-to-date modelling of the Lyman-$\alpha$ forest, and yield $\Delta_{\mathrm{lin}}^2 = 0.267^{+0.018}_{-0.023}$ and $n_{\mathrm{lin}} = -2.288^{+0.020}_{-0.020}$, with correlation coefficient $0.4$.

While both analyses exhibit a tension with $\Lambda$CDM, they correspond to different types of deviations: the Chabanier {\it et al.} likelihood is primarily driven by a change in the local slope of the spectrum at the eBOSS scale, whereas the Fernandez {\it et al.} analysis is dominated by a shift in amplitude, with a tilt closer to the CMB-preferred value.

Allowing for running and running-of-the-running induces correlated shifts in $(\Delta^2_{\rm lin}, n_{\rm lin})$ that can partially alleviate the tension with the Chabanier {\it et al.} likelihood, but cannot reach the region preferred by the Fernandez {\it et al.} analysis. This reflects the limited flexibility of the Taylor-expanded parametrisation, which only captures smooth, correlated distortions of the primordial spectrum.

Fig.~\ref{fig:corner_norunrun} shows the constraints obtained when allowing only the running of the spectral index. Fig.~\ref{fig:corner_nrunrun} extends to the running-of-the-running, $\beta_s$, showing 68 \% and 95 \% credible regions from \textit{Planck}+ACT DR6 (yellow), and +SPT-3G (blue). Upon including eBOSS Ly-$\alpha$ data, the credible regions obtained using the Chabanier {\it et al.} (purple)~\cite{eBOSS:2018qyj} and Fernandez {\it et al.} (teal)~\cite{Fernandez:2023grg} likelihoods are shown for comparison. In both cases, adding eBOSS data tightens the bounds, but it is remarkable that the corresponding favoured regions for the spectral index, its running and running-of-the-running are fairly different. The Chabanier {\it et al} likelihood shift both $n_s$ and $\alpha_s$; the data disfavours $\alpha_s=0$ at $> 2\sigma$, already hinting at higher-order scale dependence. The inclusion of eBOSS sharply contracts the contours and drives them along the $\alpha_s$–$\beta_s$ toward more negative values, with $\beta_s=0$ disfavoured at about $2\sigma$. The corresponding shift of $n_s$ toward larger values suggests a slightly bluer tilt once small-scale Ly-$\alpha$ data are included. 
On the other hand, the Fernandez {\it et al.} likelihood appears consistent with $\alpha_s=\beta_s=0$ in the corner plots, despite favouring a region in the $(\Delta^2_{\rm lin}, n_{\rm lin})$ plane that is clearly offset from the CMB contours (see Fig.~\ref{fig:nlin}). This does not indicate that zero running is preferred. Rather, it reflects the limited flexibility of the Taylor-expanded parametrisation: the type of deviation favoured by likelihood in Ref.~\cite{Fernandez:2023grg} cannot be reproduced by smooth variations controlled by $(\alpha_s,\beta_s)$. As a result, the fit remains close to the $\Lambda$CDM region in the $(n_s,\alpha_s,\beta_s)$ space. In practice, reproducing this behaviour would require either additional parameters in the running expansion or a sharper feature in the primordial power spectrum.

Since our approach is able to reconcile CMB measurements primarily with the likelihood from Ref.~\cite{eBOSS:2018qyj}, we will use the latter as an illustrative reference case in what follows. Although more recent analyses rely on improved modelling~\cite{Fernandez:2023grg, Walther:2024tcj}, this likelihood provides a concrete realisation of a slope-driven small-scale suppression that can be consistently mapped onto correlated variations of $(\alpha_s,\beta_s)$ within the Taylor-expanded framework.

Using posteriors in the $(\Delta_{\rm lin}^2,n_{\rm lin})$ plane (Fig.~\ref{fig:nlin}), we quantify the tension between CMB data and the eBOSS Lyman-$\alpha$ constraint of Chabanier {\it et al.} with the {\tt tensiometer} package~\cite{Raveri:2021wfz} (see Supplemental Material for details). The combined ACT+P+SPT data increase the CMB–eBOSS discrepancy beyond the previously reported $4.9\,\sigma$ for \Planck\ alone~\cite{Rogers:2023upm}. Allowing for running reduces this to $\sim 3$--$3.3\sigma$ (vs. $\sim 1\sigma$ with \Planck\ only), and including $\beta_s$ further lowers it to $\sim 2.8$--$3\sigma$. 
In Fig.~\ref{fig:corner_nrunrun}, this decreasing tension manifests as competing preferences: \Planck\ yields a mildly negative central value of $\alpha_s$, and ACT DR6 favours $\alpha_s>0$, whereas the eBOSS likelihood requires a strong suppression of small-scale power, corresponding to $\alpha_s<0$. Consequently, CMB data collectively prefer $\alpha_s>0$ and $\beta_s>0$, while the eBOSS constraint drives both towards negative values. Allowing for $\beta_s<0$ partially alleviates the discrepancy by steepening the small-scale suppression, but does not recover the positive running preferred by the CMB.

\begin{figure}
    \includegraphics[width=0.92\linewidth]{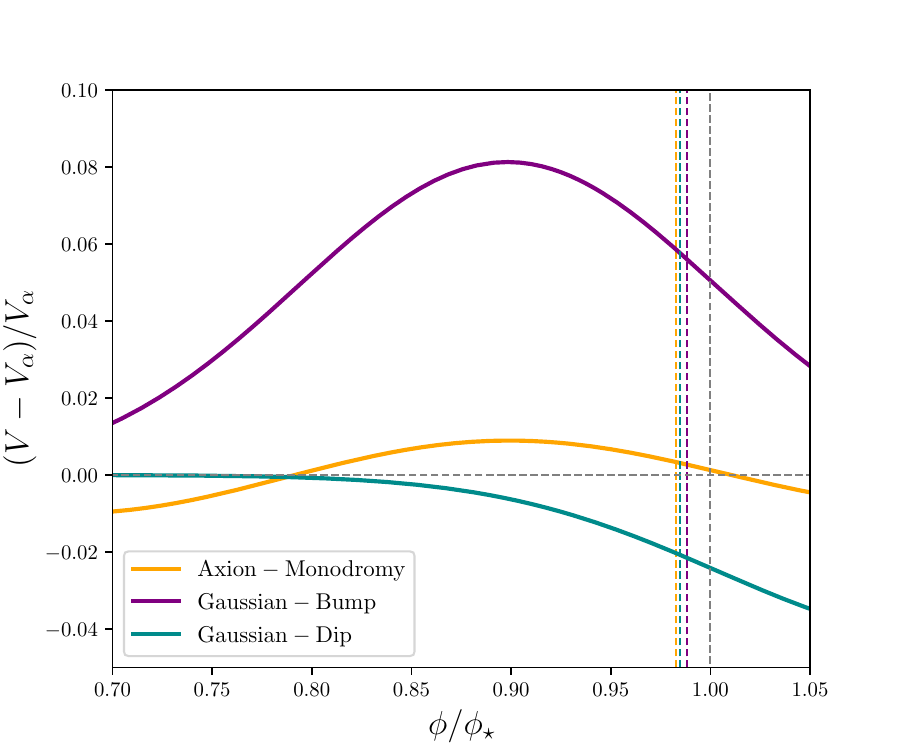}
    \caption{\footnotesize Relative modulation of the potential $(V-V_\alpha)/V_\alpha$, as compared to the baseline model $V_\alpha\equiv V_0|\phi/M_p|^\alpha$, for each set of best-fit parameters $(V_0, \alpha, \ldots)$ listed in TABLE~\ref{tab:bestfits_main_new}. Vertical, coloured, dashed lines denote the field value corresponding to where perturbations exit the horizon at the eBOSS scale.}
    \label{fig:feature_shape}
\end{figure}

\begin{table*}[htb]
  \centering
  \scriptsize
  \begin{tabular*}{\textwidth}{@{\extracolsep{\fill}} ll *{9}{c} c c c @{}}
    \toprule
    Dataset & Model
      & $\log_{10}V_0$
      & $\alpha$
      & $\log_{10}A$
      & $\phi_0\,[M_{\rm Pl}]$
      & $\sigma\,[M_{\rm Pl}]$
      & $\gamma$
      & $\log_{10}f$
      & $p_f$
      & $\rm{ln} \mathcal{L}_{\rm KDE}$
      & $\rm{ln} \pi_r$
      & $\rm{ln} \mathcal{L}$ \\
    \midrule
    ACT+P
      & Powerlaw
        & $-9.3565$
        & $0.3935$
        & --
        & --
        & --
        & --
        & --
        & --
        & $32.032$
        & $2.835$
        & $34.867$ \\
      & Powerlaw+Bump
        & $-9.5142$
        & $0.1698$
        & $-3.8106$
        & $-4.287$
        & $0.161$
        & --
        & --
        & --
        & $34.024$
        & $3.603$
        & $37.626$ \\
      & Powerlaw+Dip
        & $-9.5350$
        & $0.1951$
        & $-2.3266$
        & $-5.594$
        & $0.599$
        & --
        & --
        & --
        & $34.024$
        & $3.603$
        & $37.626$ \\
      & Axion--Monodromy
        & $-9.5085$
        & $0.1662$
        & $-4.2186$
        & --
        & --
        & $7.416$
        & $-1.053$
        & $0.0053$
        & $34.051$
        & $3.603$
        & $37.654$ \\
    \midrule
    +SPT
      & Powerlaw
        & $-9.3829$
        & $0.3067$
        & --
        & --
        & --
        & --
        & --
        & --
        & $31.730$
        & $3.334$
        & $35.065$ \\
      & Powerlaw+Bump
        & $-9.4408$
        & $0.0826$
        & $-1.5910$
        & $-4.096$
        & $0.938$
        & --
        & --
        & --
        & $35.099$
        & $3.601$
        & $38.699$ \\
      & Powerlaw+Dip
        & $-9.3917$
        & $0.0316$
        & $-1.3712$
        & $-1.071$
        & $0.988$
        & --
        & --
        & --
        & $35.099$
        & $3.601$
        & $38.699$ \\
      & Axion--Monodromy
        & $-9.5012$
        & $0.1566$
        & $-3.2910$
        & --
        & --
        & $0.492$
        & $0.077$
        & $0.829$
        & $35.099$
        & $3.603$
        & $38.702$ \\
    \midrule
    +eBOSS
      & Powerlaw
        & $-10.2659$
        & $1.6448$
        & --
        & --
        & --
        & --
        & --
        & --
        & $6.859$
        & $-19.494$
        & $-12.636$ \\
      & Powerlaw+Bump
        & $-10.2095$
        & $0.8137$
        & $-1.0900$
        & $-8.282$
        & $0.964$
        & --
        & --
        & --
        & $37.305$
        & $3.603$
        & $40.908$ \\
      & Powerlaw+Dip
        & $-9.8199$
        & $0.5068$
        & $-1.3842$
        & $-7.776$
        & $0.774$
        & --
        & --
        & --
        & $37.305$
        & $3.603$
        & $40.908$ \\
      & Axion--Monodromy
        & $-9.6943$
        & $0.3672$
        & $-1.7747$
        & --
        & --
        & $24.70$
        & $-0.372$
        & $-0.0060$
        & $37.305$
        & $3.603$
        & $40.908$ \\
    \bottomrule
  \end{tabular*}
  \scriptsize
  \caption{\label{tab:bestfits_main_new}\footnotesize
  Best-fit parameters, KDE log–likelihood ($\ln\mathcal{L}_{\rm KDE}$), and tensor prior contribution ($\ln\pi_r$) for each dataset/model/feature combination from the updated {\tt PIPE} runs. The total posterior is $\ln\mathcal{L}=\ln\mathcal{L}_{\rm KDE}+\ln\pi_r$; the Taylor penalty is zero for all listed best fits.}
\end{table*}

\begin{table*}[htb]
  \centering
  \scriptsize
  \begin{tabular*}{\textwidth}{@{\extracolsep{\fill}} ll *{5}{c}@{}}
    \toprule
    Dataset & Model
      & $A_s$
      & $r$
      & $n_s$
      & $\alpha_s$
      & $\beta_s$ \\
    \midrule
    ACT+P 
      & Powerlaw
      & $2.1367\times10^{-9}$
      & $0.0291$
      & $0.9779$
      & $-4.09\times10^{-4}$
      & $-8.81\times10^{-6}$ \\
    & Powerlaw+Gaussian-Bump
      & $2.1109\times10^{-9}$
      & $0.0125$
      & $0.9680$
      & $-4.15\times10^{-4}$
      & $-6.46\times10^{-6}$ \\
    & Powerlaw+Gaussian-Dip
      & $2.1110\times10^{-9}$
      & $0.0125$
      & $0.9680$
      & $-4.12\times10^{-4}$
      & $-6.39\times10^{-6}$ \\
    & Axion--Monodromy
      & $2.1235\times10^{-9}$
      & $0.0125$
      & $0.9713$
      & $+3.70\times10^{-3}$
      & $-5.13\times10^{-4}$ \\
    \midrule
    ACT+P+SPT 
      & Powerlaw
      & $2.1092\times10^{-9}$
      & $0.0227$
      & $0.9787$
      & $-3.94\times10^{-4}$
      & $-8.27\times10^{-6}$ \\
    & Powerlaw+Gaussian-Bump
      & $2.1209\times10^{-9}$
      & $0.0130$
      & $0.9681$
      & $+2.94\times10^{-3}$
      & $-2.92\times10^{-4}$ \\
    & Powerlaw+Gaussian-Dip
      & $2.1209\times10^{-9}$
      & $0.0130$
      & $0.9681$
      & $+2.96\times10^{-3}$
      & $-2.96\times10^{-4}$ \\
    & Axion--Monodromy
      & $2.1209\times10^{-9}$
      & $0.0125$
      & $0.9681$
      & $+2.96\times10^{-3}$
      & $-2.95\times10^{-4}$ \\
    \midrule
    ACT+P+SPT+eBOSS 
      & Powerlaw
      & $2.1547\times10^{-9}$
      & $0.1210$
      & $0.9665$
      & $-6.16\times10^{-4}$
      & $-1.65\times10^{-5}$ \\
    & Powerlaw+Gaussian-Bump
      & $2.1293\times10^{-9}$
      & $0.0125$
      & $0.9693$
      & $-8.66\times10^{-3}$
      & $-2.56\times10^{-3}$ \\
    & Powerlaw+Gaussian-Dip
      & $2.1293\times10^{-9}$
      & $0.0125$
      & $0.9693$
      & $-8.66\times10^{-3}$
      & $-2.56\times10^{-3}$ \\
    & Axion--Monodromy
      & $2.1293\times10^{-9}$
      & $0.0125$
      & $0.9693$
      & $-8.66\times10^{-3}$
      & $-2.56\times10^{-3}$ \\
    \bottomrule
  \end{tabular*}
  \caption{\footnotesize Verified spectral parameters from the best-fit potentials. 
  All quantities are evaluated at the pivot scale $k_*=0.05\,{\rm Mpc}^{-1}$.}
  \label{tab:bestfits_spectra}
\end{table*}

\section{Implications for Inflation}
Recent constraints favouring higher values of the scalar spectral index have revived interest in inflationary scenarios beyond canonical plateau potentials. Models such as exponential $\alpha$-attractors~\cite{Kallosh:2013yoa, Kallosh:2014rga,Kallosh:2013maa}, Starobinsky~\cite{Starobinsky:1980te}, and Higgs inflation~\cite{Bezrukov:2007ep} typically struggle to accommodate $ n_s \gtrsim 0.97 $~\cite{Ferreira:2025lrd, Drees:2025ngb, Haque:2025uis, Zharov:2025zjg, Ye:2025idn}. This trend motivates the exploration of alternatives that favour potentials approaching the plateau polynomially rather than exponentially~\cite{Ferreira:2025lrd}. We therefore consider several single-field potentials, beginning with a monomial baseline $V_\alpha = V_0 | \phi/M_p|^\alpha$, where $\alpha$ controls the slope and $\alpha < 1$ is preferred by recent data~\cite{ACT:2025tim}. We then introduce localised or oscillatory deviations to capture transient dynamics.
\paragraph*{Gaussian bumps and dips.}
Transient features---{\em e.g.} from localized particle production or modulations~\cite{Barnaby:2009dd,PhysRevD.82.106009,Chung:1999ve,Fumagalli:2020nvq}---can modify the dynamics of $\dot{\phi}$, impact CMB observables, or even trigger phases of ultra-slow roll leading to primordial black hole formation~\cite{Mishra:2019pzq}. We model such effects by modulating the baseline potential with
\begin{equation}
V = V_\alpha \left[ 1 \pm A \exp\left( -\frac{(\phi - \phi_0)^2}{2 \sigma^2} \right) \right],
\end{equation}
where the $+$ sign produces a bump and the $-$ a dip. The parameters $A$, $\phi_0$, and $\sigma$ set the amplitude, position, and width, controlling the induced running and feature scale.

\noindent
\paragraph*{Axion monodromy.}
Inspired by axion-like inflation in string theory~\cite{Flauger:2009ab,Flauger:2014ana}, monodromy potentials also include periodic modulations on a monomial background:
\begin{equation}
V = V_0 \left[\left| \frac{\phi}{M_p} \right|^\alpha + A \cos\left(\gamma + \frac{M_p}{f}\left|\frac{\phi}{M_p}\right|^{1+p_f} \right)\right],
\end{equation}
with amplitude $A$, phase $\gamma$, axion decay constant $f$, and index $p_f$. These oscillations naturally generate small-scale running and resonant signatures in the PPS.

\paragraph*{Inflation Observables.}
For each benchmark model and parameter set, we computed the PPS amplitude {$A_s$}, spectral index $n_s$, tensor-to-scalar ratio $r$, and scale dependence parameters $\alpha_s$ and $\beta_s$ by integrating the background equations numerically. The slow-roll parameters were evaluated analytically from the potential and its derivatives at horizon exit, to leading order in the slow-roll expansion~\cite{Auclair:2022yxs}. The PPS was then reconstructed on all relevant scales using the slow-roll approximation. We restrict to parameter choices for which the Taylor expansion of Eq.~\eqref{eq:running} provides a reliable description of the primordial power spectrum, requiring percent-level agreement with the full numerical spectrum at the eBOSS scale (see Supplemental Material for details).
\begin{table}[htb]
  \centering
  \begin{tabular*}{0.48\textwidth}{@{\extracolsep{\fill}} llcc@{}}
    \toprule
    Dataset & Model
      & $\Delta \chi^2$                  
      & $\Delta$AIC     
      \\
    \midrule
    ACT+P 
      & Powerlaw+Bump   
      & $-7.27$   
      & $-1.27$      
      \\
      & Powerlaw+Dip  
      & $-7.27$   
      & $-1.27$
      \\
      & Axion--Monodromy
      & $-7.27$   
      & $0.73$           
      \\
    \midrule
    +SPT 
      & Powerlaw+Bump   
      & $-5.52$   
      & $0.48$     
      \\
      & Powerlaw+Dip 
      & $-5.52$   
      & $0.48$ 
      \\
      & Axion--Monodromy 
      & $-5.57$   
      & $2.43$          
      \\
    \midrule
    +eBOSS 
      & Powerlaw+Bump 
      & $-107.09$   
      & $-101.09$ 
      \\
      & Powerlaw+Dip 
      & $-107.09$   
      & $-101.09$ 
      \\
      & Axion--Monodromy 
      & $-107.09$   
      & $-99.09$ 
      \\
    \bottomrule
  \end{tabular*}
  \scriptsize
  \caption{\label{tab:signi}\footnotesize
  Statistical deviation of the best-fit feature and axion--monodromy models
  relative to the baseline power-law potential for each dataset combination. 
  Negative $\Delta$AIC values indicate statistical preference over the baseline model. 
  The likelihood difference is defined as $\Delta\chi^2=-2\,(\ln\mathcal{L}_{\rm model}-\ln\mathcal{L}_{\rm powerlaw})$.}
\end{table}
\paragraph*{Likelihood.}
To compare model predictions with CMB and LSS data, we build a custom likelihood over the effective PPS parameters \(\{A_s, n_s, \alpha_s, \beta_s\}\). Posterior samples from the MCMC chains are processed with {\tt GetDist}~\cite{Lewis:2019xzd} and approximated by a multidimensional Gaussian kernel density estimator (KDE), defining a smooth empirical likelihood $\ln\mathcal{L}_{\rm KDE} (A_s, n_s, \alpha_s, \beta_s)$.  These procedures are implemented in our public code {\tt PIPE} (Potential Inflation Posterior Emulator), which provides a fast surrogate for the full CMB and eBOSS likelihoods (see Supplemental Material). 

Although the analysis focuses on the scalar spectrum, {\tt PIPE} also enforces consistency with current bounds on the tensor-to-scalar ratio by adding a BK18-based prior $\rm{ln}\pi_r$~\cite{Tristram:2021tvh} to the log-likelihood.

\paragraph*{Best-Fit Search.}
For each dataset, we maximized the {\tt PIPE} log-posterior \(\ln\mathcal{P}\equiv\ln\mathcal{L}_{\rm KDE}+\ln\pi_r\) under broad, uninformative priors on the model parameters. The optimisation combined a Genetic Algorithm ({\tt evortran}~\cite{Biekotter:2025gkp}) with a final Nelder–Mead refinement. {For every dataset–model–feature combination we repeated the GA with different random seeds and retained the best posterior value.}  This procedure yields robust potential constraints and enables consistency checks across the P–ACT, P–ACT+SPT, and P–ACT+SPT+eBOSS combinations.

\begin{figure*}
    \centering
    \includegraphics[width=0.9\linewidth]{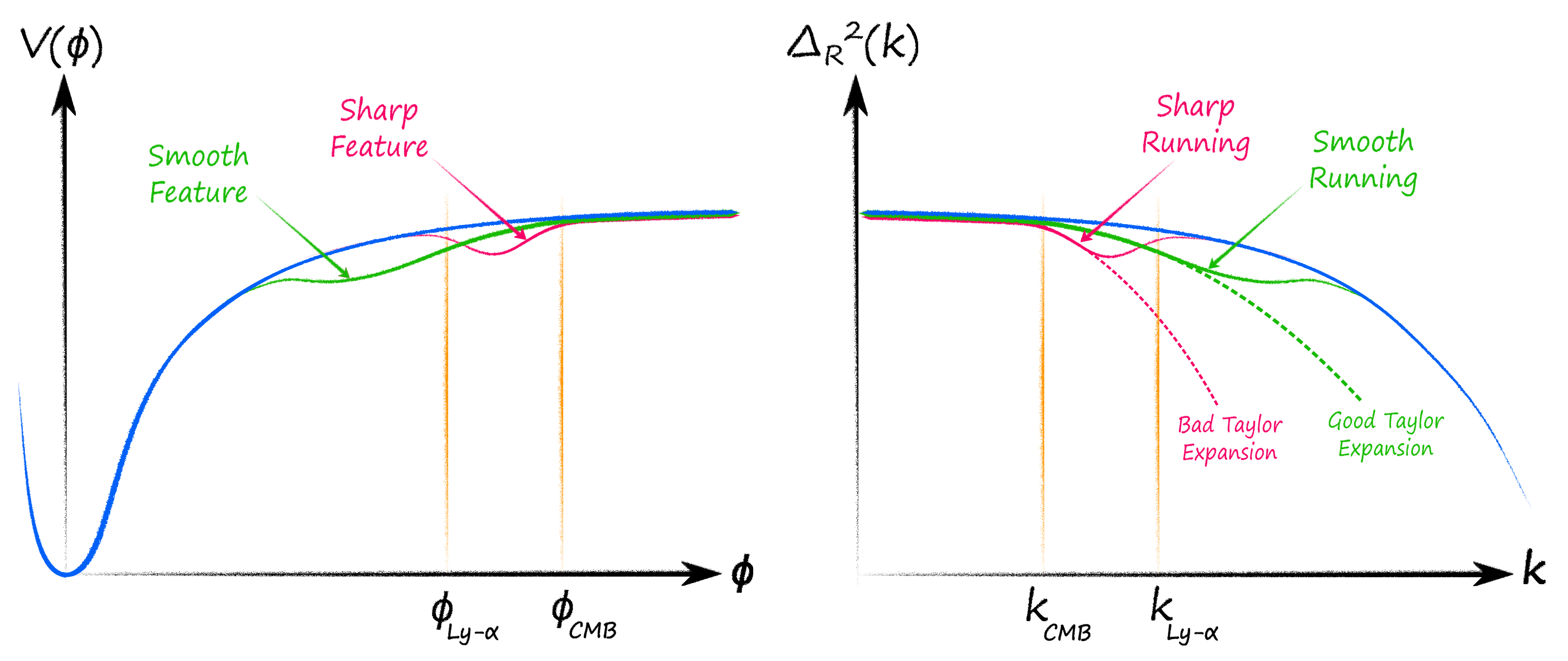}
    \caption{\footnotesize Schematic inflaton potential (left panel) featuring both a smooth feature, which can be captured by the parametrisation used in this paper, and a sharp feature, which lies beyond its range of validity. The corresponding primordial power spectrum is shown in the right panel. Smooth enough features lead to a smooth running that can be well approximated by the Taylor expansion of Eq.~\eqref{eq:running}, whereas sharper features can induce more pronounced variations that cannot be fully captured within this framework and are left for future work.} 
    \label{fig:sketch}
\end{figure*}
\paragraph*{Results.}
The best-fit potentials and spectral parameters are listed in Tables~\ref{tab:bestfits_main_new}–\ref{tab:bestfits_spectra}.  
Because multiple parameter combinations yield nearly identical effective spectra $(A_s,n_s,\alpha_s,\beta_s)$, the KDE posterior exhibits flat regions with equivalent maxima; we therefore quote a representative best-fit per model and dataset.  
Likelihood–ratio tests relative to the baseline power law give the $\Delta\chi^2$ and $\Delta$AIC values in Table~\ref{tab:signi}. For CMB data alone (ACT+P and ACT+P+SPT), the extensions provide only mild improvements, with $|\Delta\chi^2|\simeq5$–7 and small $|\Delta{\rm AIC}|\lesssim2$. When eBOSS data are included, the power-law potential becomes strongly disfavoured, while all featureful and axion–monodromy models achieve comparable maximum posteriors around $\ln\mathcal{L}\simeq40.9$. This degeneracy reflects the fact that different potentials can produce nearly identical primordial spectra over the range of scales probed. As illustrated in Fig.~\ref{fig:feature_shape}, the corresponding best-fit features induce different localised modulations of the potential around the field value associated with the eBOSS scale, yet lead to equivalent effective parameters $(A_s,n_s,\alpha_s,\beta_s)$. Within the Taylor-expanded parametrisation, these scenarios are therefore observationally indistinguishable.

\section{Summary and Discussion}
In this work, we have investigated the scale dependence of the primordial power spectrum (PPS) by combining CMB measurements from \Planck, ACT DR6, and SPT-3G with Lyman-$\alpha$ forest constraints from eBOSS. Extending the analysis beyond the scales probed by the CMB alone allows us to test departures from a simple power-law PPS and to assess their interpretation within inflationary models.

Using a Taylor expansion of $\ln \Delta^2_{\mathcal R}(k)$ around a pivot scale, we parametrised deviations from scale invariance in terms of the running $\alpha_s$ and its running $\beta_s$. We showed that, within this framework, correlated variations of $(\alpha_s,\beta_s)$ shift the CMB-preferred region in the $(\Delta_{\rm lin}^2,n_{\rm lin})$ plane along a well-defined direction. This direction aligns with the type of deviation associated with the eBOSS likelihood of Chabanier \textit{et al.}, which is characterised by a suppression of small-scale power driven primarily by a change in the local spectral slope.

When adopting this likelihood as a benchmark, we find that allowing for running partially alleviates the tension between CMB and Lyman-$\alpha$ data, reducing it from $\sim 5\sigma$ to $\sim 3\sigma$, and further to $\sim 2.8$–$3\sigma$ when including $\beta_s$. This improvement reflects the ability of the Taylor-expanded parametrisation to capture smooth, slope-driven distortions of the PPS. However, the tension is not fully resolved, as the CMB data favour mildly positive running while the eBOSS likelihood requires a strong suppression at small scales, corresponding to negative $\alpha_s$ and $\beta_s$.

A key result of this work is that not all Lyman-$\alpha$ inferences can be mapped onto this parametrisation. In particular, more recent analyses, such as those based on the \texttt{PRIYA} simulations, favour deviations that are predominantly amplitude-driven rather than slope-driven. These cannot be reproduced within the limited flexibility of the Taylor expansion, and therefore do not translate into a preference for non-zero $(\alpha_s,\beta_s)$ in parameter space. This highlights that the running parametrisation probes only a restricted class of smooth deformations of the PPS. Indeed, as it is represented in Fig.~\ref{fig:sketch}, only inflaton potentials with smooth enough features lead to primordial power spectra that can accurately be approximated by the Taylor expansion of Eq.~\eqref{eq:running}. In the contrary case of sharp features, leading to more pronounced variations of the primordial power spectrum, one would need either to push the Taylor expansion to higher and higher orders, or to perform statistical analyses at the level of the potential parameters directly while solving the Mukhanov-Sasaki equations for each set of parameters consistently. Doing so would go beyond the scope of this paper, which we leave for future work.

Motivated by these findings, we explored the inflationary origin of such deviations by reconstructing the PPS from explicit single-field potentials. We considered a monomial baseline supplemented by localised features (Gaussian bumps and dips) and oscillatory modulations (axion monodromy). We found that while CMB data alone mildly favour featureful extensions over a pure power-law potential, the statistical significance remains limited. In contrast, when including the eBOSS (Chabanier \textit{et al.}) likelihood, the power-law potential becomes strongly disfavoured, and models with features or oscillations provide a substantially better fit.

Despite this improvement, we observe a pronounced degeneracy at the level of inflationary potentials: different microphysical realisations—bumps, dips, or oscillatory modulations—lead to nearly identical effective parameters $(A_s,n_s,\alpha_s,\beta_s)$ and therefore indistinguishable observational signatures within the Taylor-expanded framework.

These results should be interpreted with care. The eBOSS likelihood adopted here does not incorporate the most recent modelling developments, and newer analyses, including DESI, are consistent with negligible running within current uncertainties. Our findings therefore do not constitute evidence for a departure from $\Lambda$CDM, but rather illustrate how a class of small-scale suppressions—--if confirmed by future measurements--—could be consistently mapped onto primordial parameters and inflationary dynamics. 

More broadly, this work clarifies the connection between late-time observables and inflationary physics in the presence of scale-dependent features, and upcoming Lyman-$\alpha$ measurements will determine whether the hints of small-scale structure suppression  seen with eBOSS persist. If confirmed, they would provide a rare window into inflationary dynamics beyond the slow-roll paradigm; if not, they will further reinforce the robustness of the minimal $\Lambda$CDM description across an extended range of scales.

\vspace{-5pt}\section*{Acknowledgements}
The authors would like to thank Hengameh Bagherian,  Jo Dunkley, and Mudit Jain for helpful discussions throughout this project, as well as Doddy Marsh and Keir Rogers for valuable comments on our manuscript.  
They are also grateful to Thomas Biekötter for his assistance with the \texttt{evortran} code.  
This work was supported by the STFC under UKRI grant ST/X000753/1.  
LH also thanks the Institute for Particle Physics Phenomenology at Durham University for access to its high-performance computing facilities.

\clearpage

\appendix

\section{Validity of the eBOSS Likelihood}
\begin{figure}
    \centering
    \includegraphics[width=\linewidth]{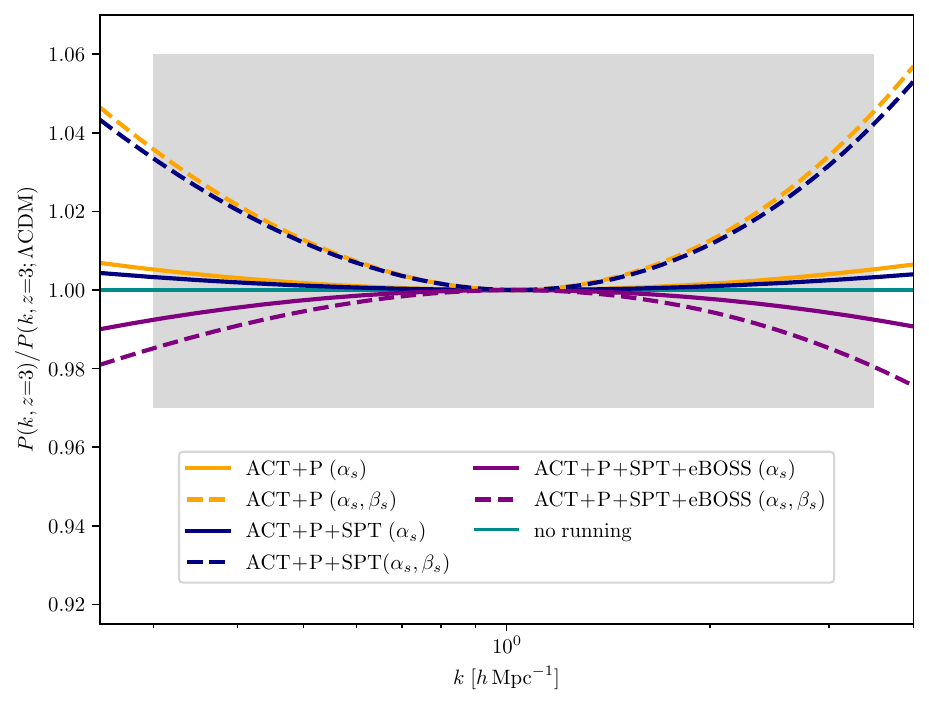}
    \caption{\label{fig:compression}\footnotesize Linear matter power spectrum, rescaled both in amplitude and scale compared to the $\Lambda$CDM best fit from \Planck\cite{Planck:2018vyg} at the eBOSS scale, and compared to the eBOSS range of frequencies and credible region (grey-shaded area) as defined in Ref.~\cite{Chabanier:2019eai}.}
\end{figure}

To assess the validity of the compressed eBOSS likelihood we used in the analysis, we checked that the linear matter power spectrum obtained remains sufficiently close to a power-law, within the $1\sigma$ validity band of the eBOSS measurement. To do so, we used a similar approach as in Ref.~\cite{Rogers:2023upm}, by taking the best-fit linear matter power
spectra for each dataset combination, and rescaling them such that they
have the same values of $\Delta^2_{\rm lin}$, $n_{\rm lin}$ as the best-fit $\Lambda$CDM model from \Planck\cite{Planck:2018vyg} at the eBOSS scale. Results are depicted in FIG.~\ref{fig:compression}, where one can see that each best fit point leads to a linear matter power spectrum that is contained within the eBOSS range of frequencies and credible region, as calculated in  Ref.~\cite{Chabanier:2019eai} (grey-shaded area), meaning that the linear matter power spectrum obtained is sufficiently power-law-like for the eBOSS compression to remain valid over the appropriate range of scales.

\section{Taylor Expansion}
\label{sec:taylorexp}
When confronting power spectra obtained from slow-roll inflation potentials, we used the Taylor expansion of Eq.~\eqref{eq:running} at the pivot scale, and assumed that this expansion is valid up to the eBOSS scale $k\simeq 1 h \mathrm{Mpc^{-1}}$, so we can use the constraints on $\{A_s, n_s, \alpha_s,\beta_s\}$ obtained from CMB and Ly-$\alpha$ likelihoods to constrain potential parameters. To assess the validity of this assumption, we calculated the PPS using the slow-roll approximation numerically and compared its value at the eBOSS scale. We imposed the relative difference to be within $\delta\equiv|\Delta_{\mathcal R \mathrm{,\ Taylor}}^2-\Delta_{\mathcal R}^2|/\Delta_{\mathcal R}^2 < \delta_{\rm thresh}$, where we chose $\delta_{\rm thresh}=1\%$ so that the corresponding error made on the linear matter power spectrum at the eBOSS scale lies within the 1$\sigma$ sensitivity bound of the eBOSS measurement that is of order $\mathcal O(10)\%$ \cite{Chabanier:2019eai}. To enforce this condition while guiding the genetic algorithm to the right region of the parameter space, we added to the likelihood an extra contribution
\begin{equation}
    \Delta \mathcal L_{\mathrm{Taylor}} = \left\{\begin{matrix}
        \left(\frac{\delta-\delta_{\rm thresh}}{\delta_{\rm thresh}}\right)^p & \qquad \text{if $\delta>\delta_{\rm thresh}$}\\
        0 & \qquad \text{if $\delta<\delta_{\rm thresh}$}
    \end{matrix}\right.\,,
\end{equation}
Furthermore, because the eBOSS data spans a larger range of frequencies, we also demanded that this error does not exceed 3\% at $k_{\rm eBOSS\,,\ \max}\approx3.5 h\mathrm{Mpc^{-1}}$. 
We checked that every best fit found by the algorithm lies within the region where $\Delta \mathcal L_{\mathrm{Taylor}} = 0$, so that this guiding penalty does not contribute to the value of the likelihood. In our analysis, we used $p=2$, although our code allows for chosing arbitrary values of $p$,   $\delta_{\rm thresh}$, and the possibility to adding an additional pre-factor to the penalty.

\section{PIPE: Potential Inflation Posterior Emulator}\label{app:pipe}

\begin{figure}[h!]
  \centering
  \includegraphics[width=0.47\textwidth]{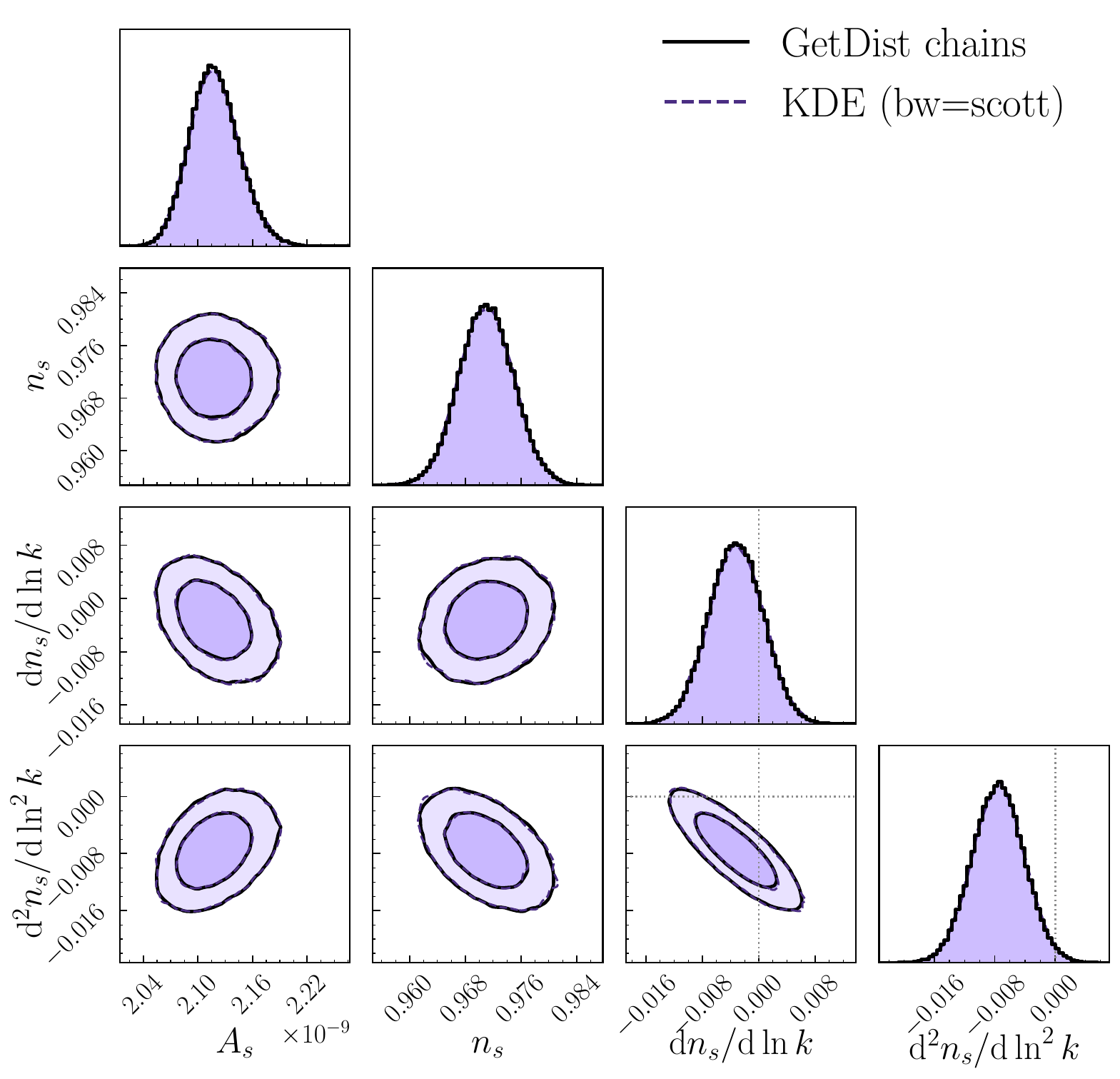}
  \caption{%
    Comparison for the P-ACT+SPT+eBOSS dataset of the 1D and 2D posterior distributions for the effective inflationary parameters 
    \(\{A_s,\,n_s,\,\alpha_s,\,\beta_s\}\) obtained directly from the \texttt{GetDist} chains (black outlines) 
    and from a Gaussian KDE resampling (purple fill) using our public \texttt{PIPE} code. The band width has been set according to the Scott's rule.
  }
  \label{fig:corner-kde-comparison}
\end{figure}

{To streamline the comparison between theoretical predictions and observational constraints, we developed a public Python package called \texttt{PIPE} (Potential Inflation Posterior Emulator), which can be found in
\begin{equation}
\href{https://gitlab.com/cosmoPipe/pipe-inflation\#}{\texttt{gitlab.com/cosmoPipe/pipe-inflation}}.
\end{equation} 
\texttt{PIPE} provides a fast, data–driven likelihood for inflationary models by emulating the joint posterior of effective spectral parameters 
\((A_s,n_s,\alpha_s,\beta_s)\) from the P-ACT, P-ACT+SPT, and P-ACT+SPT+eBOSS datasets.
The code reads in \texttt{GetDist} MCMC chains of effective parameters and constructs a smoothed empirical likelihood via a Gaussian kernel density estimator (KDE) with a bandwidth tuned for each dataset according to Scott’s rule. 
It then allows the user to specify either built–in or fully custom inflationary potentials; for each parameter point \(\theta\), 
\texttt{PIPE} evolves the background equations and returns the predicted spectral parameters 
\([A_s,r,n_s,\alpha_s,\beta_s]\) at leading order in slow roll, following Ref.~\cite{Auclair:2022yxs}.  
These predictions are directly evaluated against the KDE likelihood to obtain $\ln\mathcal{L}_{\rm KDE}$, optionally combined with an $r$-prior \(\rm{ln} \pi_r\) from BK18 via \texttt{LogL\_r.py}, 
yielding the total log-posterior \(\rm{ln}\mathcal{L}_{\rm KDE}+\rm{ln}\pi_r\).  
A Taylor-expansion penalty is also implemented (as described in \ref{sec:taylorexp}).

Internally, \texttt{PIPE} consists of:
\begin{itemize}
    \item \texttt{pipe/Inflation\_evo.py}, which computes \([A_s,r,n_s,\alpha_s,\beta_s]\) for supported and user–provided potentials.
    \item \texttt{pipe/LogL\_KDE.py}, which constructs the KDE likelihood directly from \texttt{GetDist} chains.
    \item \texttt{pipe/LogL\_r.py}, which configures the BK18 $r$-prior~\cite{Tristram:2021tvh}.
    \item \texttt{example/example.py}, an end–to–end demonstration showing how to use the built–in \texttt{Powerlaw} model with the same custom potential, and compute \(\rm{ln} \mathcal{L}_{\rm KDE}\) all the contributions to the likelihood.

\end{itemize}

Built–in templates include \texttt{Powerlaw} (with optional Gaussian-dip or bump features) and \texttt{Monodromy}; new models can be added by providing the potential and its first four derivatives. 

Fig.~\ref{fig:corner-kde-comparison} shows, for the P-ACT+EPT+eBOSS dataset, a direct comparison between the raw \texttt{GetDist} chains (black outlines) and their KDE approximation (purple fill) produced with \texttt{PIPE}, illustrating how the KDE accurately reproduces the full MCMC posterior used in our analysis.}

\vfill
\pagebreak
\section{Effect on $\sigma_8$}\label{app:sigma8}
\begin{figure*}
    \centering
    \includegraphics[width=0.49\linewidth]{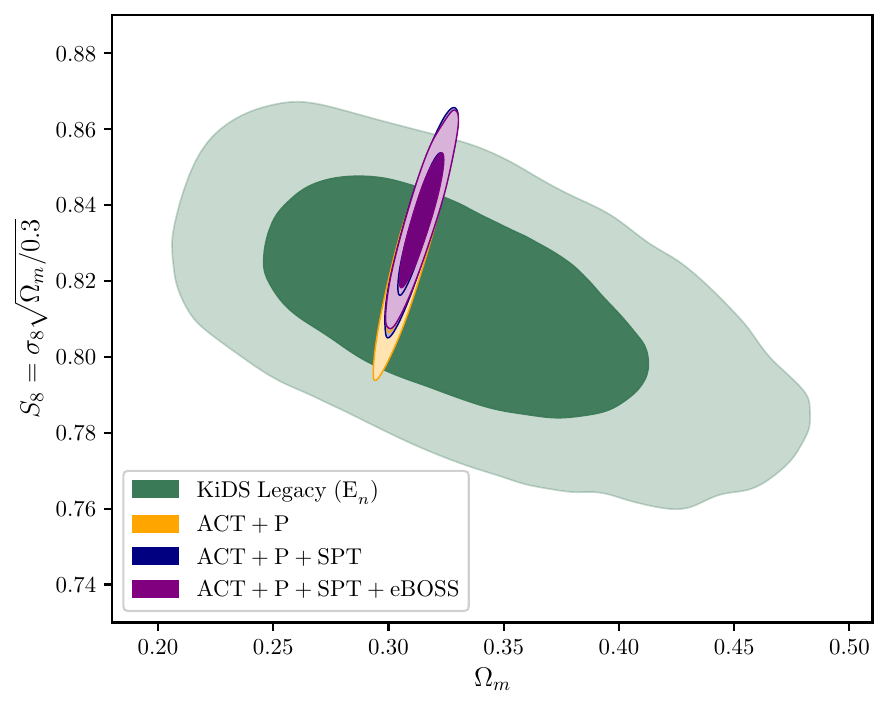}
    \includegraphics[width=0.49\linewidth]{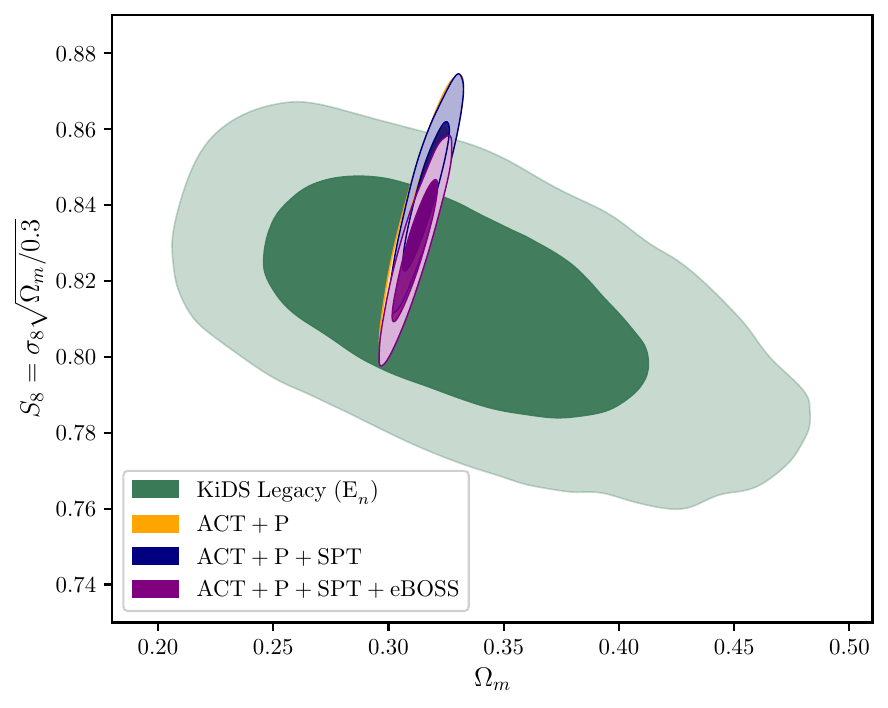}
    \caption{\footnotesize Posteriors for $\Omega_{m},S_8$ when includin the running (left panel) and running-of-the-running (right panel) of the spectral index. For comparison, green contours depict the credible region, as preferred by KiDS Legacy~\cite{Wright:2025xka}. }
    \label{fig:sigma8}
\end{figure*}

We survey here to which extent the running of the PPS's spectral index favoured by the CMB datasets considered affect the value of $\sigma_8$ and compare it to the existing constraints from KiDS Legacy~\cite{Wright:2025xka}. As one can see from FIG.~\ref{fig:sigma8}, inclusion of the running and running of the running affect very mildly the $\sigma_8$ and remains consistent with current data, with SPT asking for a slightly larger $\sigma_8$ than ACT.  The discrepancy remains between the CMB values and the other weak lensing surveys,  but the tension is not significantly altered when we include the running considered in this work.
\section{Full Cosmological Results}
\begin{figure*}
    \centering
    \includegraphics[width=\linewidth]{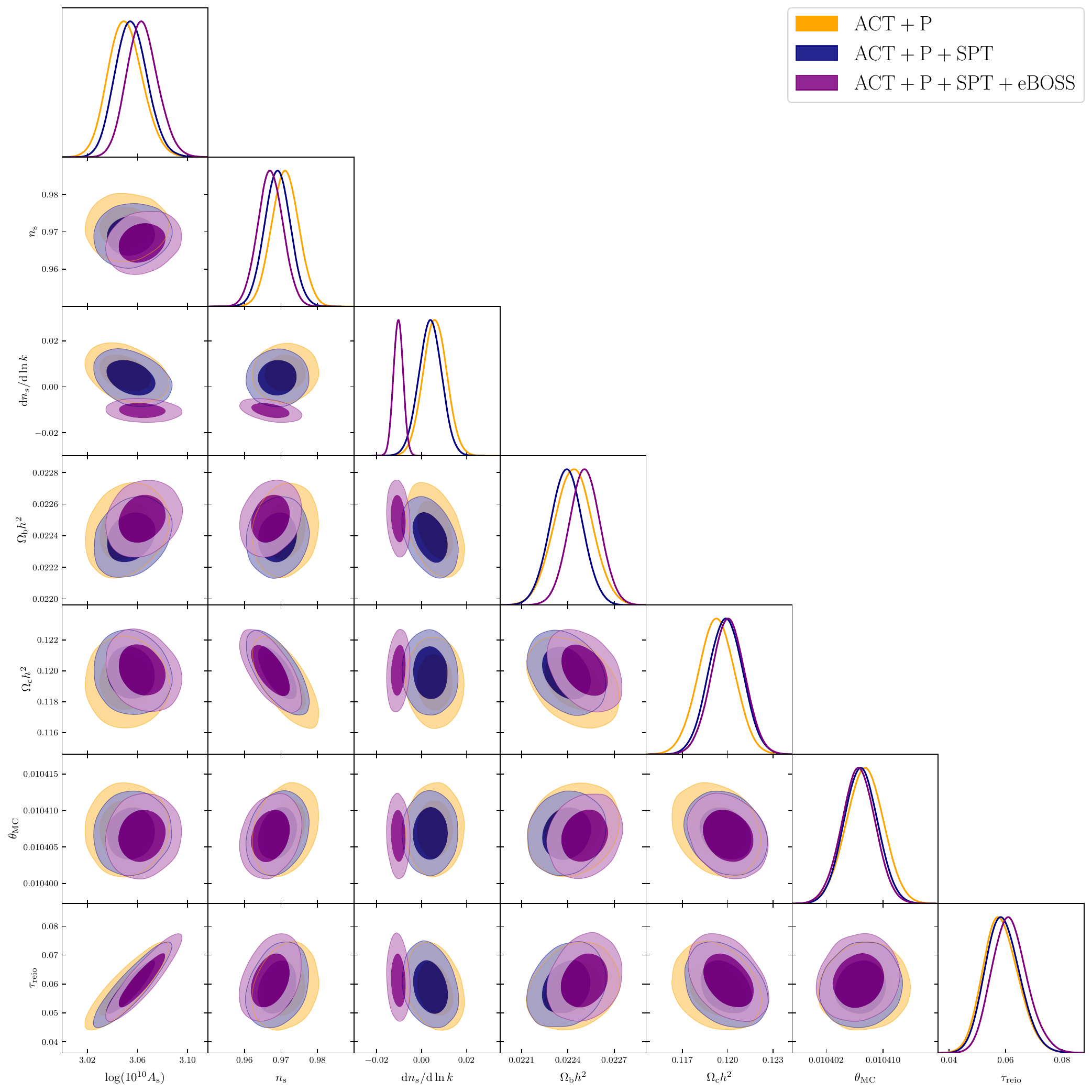}
    \caption{\label{fig:full_cosmo} \footnotesize Full cosmological 1D and 2D posteriors obtained when including the running of the spectral index $\alpha_s$.}
\end{figure*}

\begin{figure*}
    \centering
    \includegraphics[width=\linewidth]{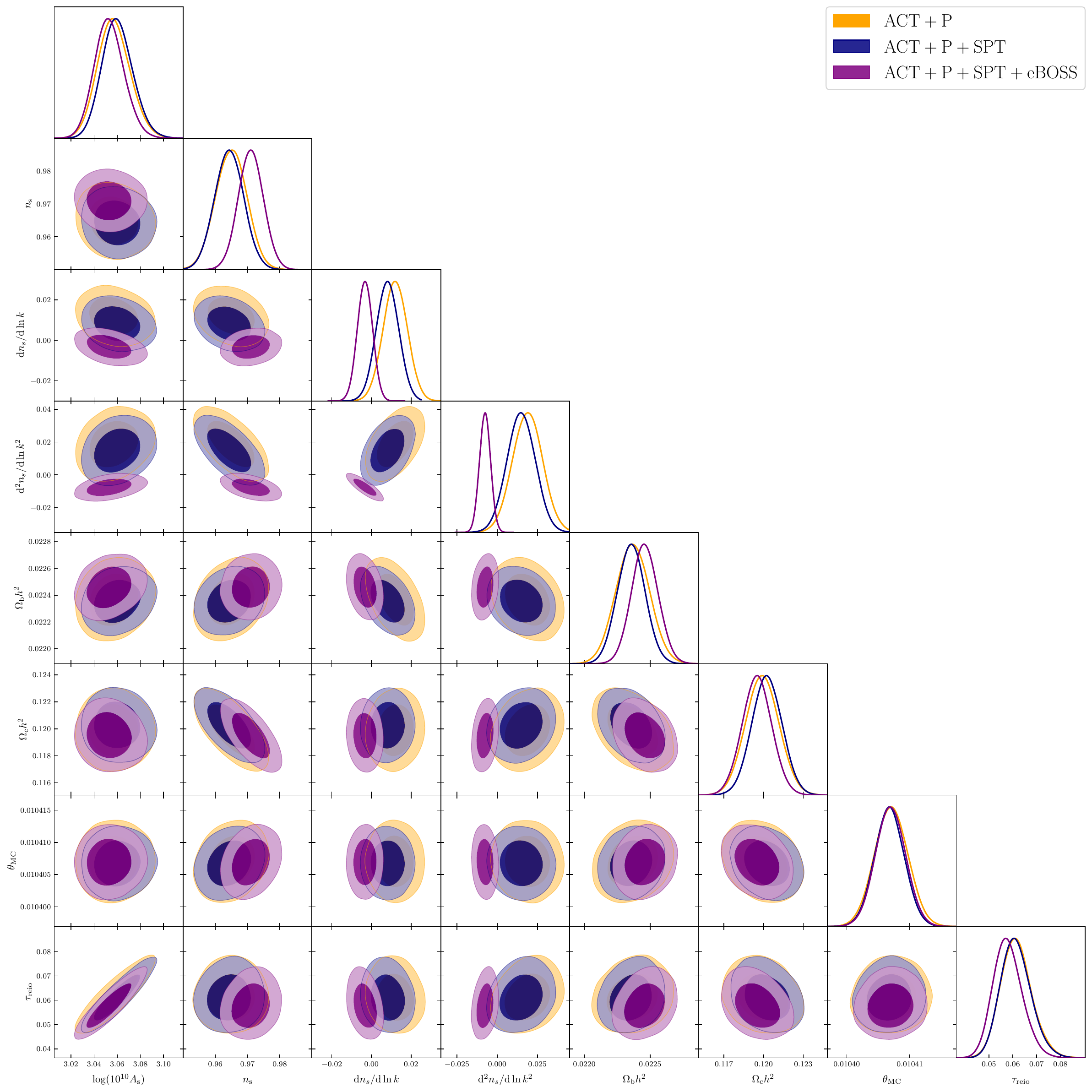}
    \caption{\label{fig:full_cosmo_runrun} \footnotesize Full cosmological 1D and 2D posteriors obtained when including the running of the spectral index $\alpha_s$, and the running-of-the-running $\beta_s$.}
\end{figure*}

We report in FIG. \ref{fig:full_cosmo} and \ref{fig:full_cosmo_runrun} the full 1D and 2D cosmological posteriors obtained when including either $\alpha_s$ or $\alpha$ and $\beta_s$ in our analysis.
\section{Primordial Power Spectrum Median Values and Statistical Tensions with $\Lambda$CDM}\label{app:tensions}
\begin{table}[h!]
\centering
\renewcommand{\arraystretch}{1.7}
\begin{tabular}{lccc}
\hline
Dataset        & $n_s$                            & $\alpha_s$                              & $\beta_s$                               \\
\hline
ACT+P           & $0.96501^{+0.00490}_{-0.00493}$ & $0.01195^{+0.00623}_{-0.00628}$         & $0.01869^{+0.00954}_{-0.00967}$         \\
+ SPT           & $0.96437^{+0.00453}_{-0.00449}$ & $0.00804^{+0.00567}_{-0.00571}$         & $0.01477^{+0.00884}_{-0.00876}$         \\
+ eBOSS         & $0.97101^{+0.00391}_{-0.00388}$ & $-0.00323^{+0.00390}_{-0.00388}$        & $-0.00755^{+0.00346}_{-0.00347}$        \\
\hline
\end{tabular}
\caption{Median values and 68\% credible intervals (16th and 84th percentiles) for $n_s$, $\alpha_s$, and $\beta_s$ in three data combinations: ACT+\Planck (ACT+P), with added SPT data (+SPT), and with both SPT and eBOSS data (+eBOSS).}
\label{tab:medians_nrunrun}
\end{table}
The median values and 68\% credible intervals (16th and 84th percentiles) for $n_s$, $\alpha_s$, and $\beta_s$ obtained from different dataset combinations are listed in TABLE~\ref{tab:medians_nrunrun}.

For the case of $\Lambda$CDM, and when including either the running alone ($\alpha_s$) or both the running and running-of-the-running ($\beta_s$) of the spectral index, we used the package {\tt tensiometer} \cite{Raveri:2021wfz} to estimate the statistical tension between CMB observatories and eBOSS measurements. This package allows to calculate the statistical tension between distributions that are not necessarily gaussian, which is exactly what is needed in our case. In each case, we tested that the result obtained is stable when varying internal settings used by {\tt tensiometer}, such as the smoothing scale used in  the estimate of the KDE shift---integrated square error (MISE), adaptive bandwidth (BALL), asyntotic MISE estimator (AMISE), or maximum bandwidth (MAX)---or the {\em boost} sampling parameter, which we varied to larger and larger values until obtaining a stable result. Our results are presented in TABLE~\ref{tab:tensions}, in which, for each entry, a subset of CMB likelihoods (row entries) is tensioned against the eBOSS likelihood from Ref.~\cite{eBOSS:2018qyj} given the model chosen (column entries). Note that, in the case of $\Lambda$CDM, {\tt tensiometer} failed at returning either a stable or finite result, due to the too small overlap between the CMB and eBOSS distributions. Some finite results returned were above 5.7 $\sigma$, which allowed us to safely claim a tension larger than $5\sigma$ both for ACT+P and ACT+P+SPT.
\begin{table}[h!]
\centering
\renewcommand{\arraystretch}{1.7}
\begin{tabular}{lccc}
\hline
Dataset &\qquad $\Lambda$CDM &\qquad + $\alpha_s$ &\qquad + ($\alpha_s$,$\beta_s$) \\
\hline
P--ACT  &\qquad $\gg 5\ \sigma$ &\qquad $3.375^{+0.002}_{-0.002}\ \sigma$ &\qquad $3.074^{+0.001}_{-0.001}\ \sigma$ \\
+SPT    &\qquad $\gg 5\ \sigma$ &\qquad $3.038^{+0.001}_{-0.001}\ \sigma$ &\qquad $2.813^{+0.001}_{-0.001}\ \sigma$ \\
\hline
\end{tabular}
\caption{Statistical tension between different CMB dataset combinations: Planck+ACT (P--ACT), with added SPT data (+SPT), and eBOSS data (+eBOSS).}
\label{tab:tensions}
\end{table}

\bibliographystyle{apsrev4-1}
\bibliography{main.bib}
\vfill
\end{document}